\documentclass{MYappolb}
\usepackage{epsfig,cite}

\def\be{\begin{equation}}
\def\ee{\end{equation}}
\def\beqn{\begin{eqnarray}}
\def\eeqn{\end{eqnarray}}
\def\no{\nonumber}
\def\ba{\begin{array}{c}}
\def\bat{\begin{array}{cc}}
\def\ea{\end{array}}
\def\bi{\begin{itemize}}
\def\ei{\end{itemize}}

\def\cO{{\cal O}}

\newcommand{\eqn}[1]{(\ref{#1})}
\newcommand{\bel}[1]{\be\label{#1}}

\def\refjl#1#2#3#4#5#6{\bibitem{#1} #2, {#3} {#4} (#5) #6.}

\begin{document}
\title{$\alpha_s$ Determination from $\tau$ Decays: Theoretical Status
\thanks{Presented at the Flavianet topical workshop on {\it Low energy constraints on extensions of
the Standard Model}, Kazimierz, Poland, 23-27 July 2009}
}
\author{Antonio Pich
\address{Departament de F\'{\i}sica Te\`{o}rica, IFIC, Universitat de Val\`{e}ncia - CSIC\\ Apt. Correus 22085, E-46071 Val\`{e}ncia, Spain}
}
\maketitle
\begin{abstract}
The total $\tau$ hadronic width can be accurately calculated using analyticity and the operator product expansion.
The result turns out to be very sensitive to the value of $\alpha_s(m_\tau^2)$,
providing a precise determination of the strong coupling constant.
The theoretical description of this observable is updated, including the recently computed $\cO(\alpha_s^4)$ contributions.
The experimental determination of $\alpha_s(m_\tau^2)$ and its actual uncertainties are discussed.
\end{abstract}

\section{Introduction}

The inclusive hadronic decay width of the $\tau$ lepton provides one of the most precise measurements
of the strong coupling \cite{BNP:92,LDP:92a,BR:88,NP:88,QCD:94,DHZ:05,QCD08}.
Moreover, the comparison of $\alpha_s(m_\tau^2)$ with
$\alpha_s$ determinations at higher energies constitutes the most accurate test of  asymptotic freedom,
successfully confirming the predicted running of the QCD coupling at the four-loop level.

The calculation of the $\cO(\alpha_s^4)$ contribution \cite{BChK:08}
has triggered a renewed theoretical interest on the $\alpha_s(m_\tau^2)$ determination, since it allows to
push the accuracy to the four-loop level \cite{BChK:08,DDMHZ:08,BJ:08,MY:08,ME:09,CF:09}.
However, the recent theoretical analyses slightly disagree on the final result,
giving rise to a range of different values for $\alpha_s(m_\tau^2)$.
The differences among these results, shown in Table~\ref{tab:alpha-pub-values},
are too large compared with the claimed $\cO(\alpha_s^4)$ accuracy and
originate in the different inputs or theoretical procedures which have been adopted.

In the following I try to clarify the reasons behind these numerical discrepancies and reassess the actual
uncertainties of the $\tau$ decay determination of $\alpha_s$. Using all present experimental and theoretical knowledge,
I derive the value
\bel{eq:alpha-result}
\alpha_s(m_\tau^2) = 0.342 \pm 0.012\, .
\ee
%

\begin{table}[tb]\centering
\begin{tabular}{|c|c|c|c|}
\hline
Reference & Method & $\delta_P$ & $\alpha_s(m_\tau^2)$  
\\ \hline
Baikov et al. 
\cite{BChK:08} & {\small CIPT, FOPT} & $0.1998 \pm 0.0043$ & $0.332\pm 0.016$
\\
Davier et al. \cite{DDMHZ:08} & {\small CIPT} & $0.2066\pm 0.0070$ & $0.344\pm 0.009$
\\
Beneke-Jamin \cite{BJ:08} & {\small BSR + FOPT} & $0.2042\pm 0.0050$ & $0.316\pm 0.006$
\\
Maltman-Yavin \cite{MY:08} & {\small PWM + CIPT} & --- & $0.321 \pm 0.013$
\\
Menke \cite{ME:09} & {\small CIPT, FOPT} 
& $0.2042\pm 0.0050$ & $0.342\: {}^{+\: 0.011}_{-\: 0.010}$
\\
Caprini-Fischer \cite{CF:09} & {\small BSR + CIPT} & $0.2042\pm 0.0050$ & $0.320\: {}^{+\: 0.011}_{-\: 0.009}$
\\ \hline
\end{tabular}
\caption{$\cO(\alpha_s^4)$ determinations of $\alpha_s(m_\tau^2)$.} 
\label{tab:alpha-pub-values}
\end{table}

\section{Theoretical framework}

The inclusive character of the total $\tau$ hadronic width renders
possible an accurate calculation of the ratio
\cite{BNP:92,LDP:92a,BR:88,NP:88,QCD:94}
\be
 R_\tau \,\equiv\, { \Gamma [\tau^- \to \nu_\tau\,\mathrm{hadrons}\, (\gamma)] \over
 \Gamma [\tau^- \to \nu_\tau e^- {\bar \nu}_e (\gamma)] } \, =\,
 R_{\tau,V} + R_{\tau,A} + R_{\tau,S}\, .
\ee
The theoretical analysis involves the two-point correlation functions for the
left-handed quark currents
$\, L^{\mu}_{ij} = \bar{\psi}_j \gamma^{\mu} (1-\gamma_5) \psi_i \; $ ($i,j=u,d,s$):
\beqn\label{eq:pi_v}
\Pi^{\mu \nu}_{ij}(q) &\equiv &
 i \int d^4x \, e^{iqx}\;
\langle 0|T(L^{\mu}_{ij}(x) L^{\nu}_{ij}(0)^\dagger)|0\rangle
\no\\ & = &
  \left( -g^{\mu\nu} q^2 + q^{\mu} q^{\nu}\right) \, \Pi_{ij}^{(1)}(q^2)
  \,  +\,   q^{\mu} q^{\nu} \, \Pi_{ij}^{(0)}(q^2) \, .
\eeqn
Using analyticity, $R_\tau$ can be written as a contour integral
in the complex $s$-plane running counter-clockwise around the circle $|s|=m_\tau^2$:
\bel{eq:circle}
 R_\tau \, =\, 6 \pi i \oint_{|s|=m_\tau^2} {ds \over m_\tau^2} \,
 \left(1 - {s \over m_\tau^2}\right)^2\,
\left[ \left(1 + 2 {s \over m_\tau^2}\right) \Pi^{(0+1)}(s)
         - 2 {s \over m_\tau^2} \Pi^{(0)}(s) \right] \, ,
\ee
where \ $\Pi^{(J)}(s) \equiv |V_{ud}|^2 \, \Pi^{(J)}_{ud}(s) + |V_{us}|^2 \, \Pi^{(J)}_{us}(s)$.
This expression requires the correlators only for
complex $s$ of order $m_\tau^2$, which is significantly larger than the scale
associated with non-perturbative effects.
Using the Operator Product Expansion (OPE)
to evaluate the contour integral, $R_\tau$
can be expressed as an expansion in powers of $1/m_\tau^2$.
%
The uncertainties associated with the use of the OPE near the
time-like axis are heavily suppressed by the presence in (\ref{eq:circle})
of a double zero at $s=m_\tau^2$ \cite{BNP:92,DDMHZ:08,CGP:08}.

The theoretical prediction for the Cabibbo-allowed decay width
can be written as \cite{BNP:92}
\begin{equation}\label{eq:Rv+a}
 R_{\tau,V+A} \, =\, N_C\, |V_{ud}|^2\, S_{\mathrm{EW}} \left\{ 1 +
 \delta_{\mathrm{P}} + \delta_{\mathrm{NP}} \right\}\, ,
\end{equation}
where $N_C=3$ is the number of quark colours
and $S_{\mathrm{EW}}=1.0201\pm 0.0003$ contains the
electroweak radiative corrections \cite{MS:88,BL:90,ER:02}.
The dominant correction ($\sim 20\%$) is the perturbative QCD
contribution in the massless-quark limit $\delta_{\mathrm{P}}$, which is already known to
$O(\alpha_s^4)$ \cite{BNP:92,BChK:08}. Quark mass effects  \cite{BNP:92,PP:99,BChK:05}
are tiny for the Cabibbo-allowed current
and amount to a negligible correction smaller than $10^{-4}$ \cite{BNP:92,BJ:08}.

Non-perturbative contributions are suppressed by six powers of the
$\tau$ mass \cite{BNP:92} and, therefore, are very small. Their
numerical size has been determined from the invariant-mass
distribution of the final hadrons in $\tau$ decay, through the study
of weighted integrals \cite{LDP:92b}
which can be calculated theoretically in the same way as $R_{\tau}$.
The predicted suppression \cite{BNP:92} of the non-perturbative
corrections has been confirmed by ALEPH \cite{ALEPH:05}, CLEO
\cite{CLEO:95} and OPAL \cite{OPAL:98}. The most recent analysis
\cite{DHZ:05} gives
\begin{equation}\label{eq:del_np}
 \delta_{\mathrm{NP}} \, =\, -0.0059\pm 0.0014 \, .
\end{equation}

The measured values of the $\tau$ lifetime and leptonic branching ratios imply
$R_\tau = 3.640\pm 0.010$ \cite{DDMHZ:08}. Subtracting the Cabibbo-suppressed
contribution $R_{\tau,S}= 0.1615 \pm 0.0040$ \cite{DDMHZ:08}, one obtains
$R_{\tau,V+A} = 3.479\pm 0.011$. Using $|V_{ud}| = 0.97418\pm 0.00027$ \cite{PDG:08}
and \eqn{eq:del_np}, the pure perturbative contribution to $R_\tau$ is determined to be:
\bel{eq:delta_P}
\delta_P = 0.2038 \pm 0.0040 \, .
\ee

\section{Perturbative contribution to $\mathbf{R_\tau}$}

In the chiral limit ($m_u=m_d=m_s=0$), the vector and axial-vector currents are conserved.
This implies  $s \,\Pi^{(0)}(s) = 0$; therefore, only the correlator
$\Pi^{(0+1)}(s)$ contributes to Eq.~(\ref{eq:circle}).
The result is more conveniently expressed in terms of the
logarithmic derivative of the two-point correlation function of the vector (axial) current,
$\Pi(s)=\frac{1}{2}\,\Pi^{(0+1)}(s)$,
which satisfies an homogeneous renormalization--group equation:
\be\label{eq:Adler}
D(Q^2)  \,\equiv\,  - Q^2 {d \over dQ^2 } \Pi(Q^2)\,
=\,  {N_C\over 12 \pi^2}\, \sum_{n=0}  K_n \left( {\alpha_s(Q^2)\over \pi}\right)^n\,  .
\ee
With the choice of renormalization scale $\mu^2=Q^2\equiv - s$ all logarithmic corrections,
proportional to powers of $\log{(-s/\mu^2)}$, have been summed into the running coupling.
The $K_n$ coefficients are known to order $\alpha_s^4$. For $n_f=3$ quark flavours, one has
\cite{BChK:08,ChKT:79,DS:79,CG:80,GKL:91,SS:91}:
$K_0  =  K_1 = 1$, $K_2  = 1.63982$, $K_3^{\overline{\mathrm{MS}}} = 6.37101$ and
$K_4^{\overline{\mathrm{MS}}}  = 49.07570$.

The perturbative component of $R_\tau$ is given by
\be\label{eq:r_k_exp}
\delta_P\, =\, \sum_{n=1}\,  K_n \, A^{(n)}(\alpha_s) \, ,
\ee
where the functions \cite{LDP:92a}
\bel{eq:a_xi}
A^{(n)}(\alpha_s) \, \equiv \, {1\over 2 \pi i}\,
\oint_{|s| = m_\tau^2} {ds \over s} \, \left({\alpha_s(-s)\over\pi}\right)^n\,
\left( 1 - 2 {s \over m_\tau^2} + 2 {s^3 \over m_\tau^6} - {s^4 \over  m_\tau^8} \right)
\ee
are contour integrals in the complex plane which only depend on the strong coupling.
Using the exact solution
(up to unknown $\beta_{n>4}$ contributions) for $\alpha_s(s)$
given by the renormalization-group $\beta$-function equation,
they can be numerically computed with a very high accuracy \cite{LDP:92a}.

%
\begin{table}[tb]\centering
\begin{tabular}{|c|c|c|c|c|c|c|}
\hline
Loops & $A^{(1)}(\alpha_s)$ &
$A^{(2)}(\alpha_s)$ & $A^{(3)}(\alpha_s)$ & $A^{(4)}(\alpha_s)$ & $A^{(5)}(\alpha_s)$ &
$\delta_P$
\\ \hline
$1$ & $0.13247$ & $0.01570$ & $0.001698$ & $0.000169$ & $0.0000154$ & $0.1773$ \\ 
$2$ & $0.13523$ & $0.01575$ & $0.001629$ & $0.000151$ & $0.0000124$ & $0.1788$ \\ 
$3$ & $0.13540$ & $0.01565$ & $0.001597$ & $0.000145$ & $0.0000115$ & $0.1784$ \\ 
4 & $0.13529$ & $0.01557$ & $0.001579$ & $0.000142$ & $0.0000112$ & $0.1779$  
\\ \hline
\end{tabular}
\caption{ Exact results
for $A^{(n)}(\alpha_s)$ ($n\le 5$) at different $\beta$-function approximations,
and corresponding values of \ $\delta_P = \sum_{n=1}^4\, K_n\, A^{(n)}(\alpha_s)$,
for $\alpha_s(m_\tau^2)/\pi=0.1$.}
\label{tab:Afun}
\end{table}
%

Table~\ref{tab:Afun} gives the numerical values
for $A^{(n)}(\alpha_s)$ ($n\le 5$) obtained at the one-, two-, three- and four-loop
approximations (i.e. $\beta_{n>1}=0$, $\beta_{n>2}=0$, $\beta_{n>3}=0$ and $\beta_{n>4}=0$,
respectively), together with the corresponding results for $\delta_P = \sum_{n=1}^4\, K_n\, A^{(n)}(\alpha_s)$,
taking $\alpha_s(m_\tau^2)/\pi=0.1$.
The perturbative convergence is very good.
The error induced by the truncation of the $\beta$ function at fourth order can be conservatively estimated
through the variation of the results at five loops, assuming $\beta_5 =\pm \beta_4^2/\beta_3 = \mp 443$,
i.e. a geometric growth of the $\beta$ function. Higher-order contributions to the Adler function $D(Q^2)$ will be taken into account
adding the fifth-order term $ K_5\, A^{(5)}(\alpha_s)$ with $K_5 = 275\pm 400$. Including, moreover, the 5-loop variation with changes
of the renormalization scale in the range $\mu^2 \in [0.5,1.5]$, one gets the final result
$\delta_P = 0.1810\pm 0.0045_{K_5}\pm 0.0013_{\beta_5}\pm 0.0013_{\mu} = 0.1810\pm 0.0049$
for $\alpha_s(m_\tau^2)/\pi=0.1$.

Adopting this very conservative procedure, the experimental value of $\delta_P$ given in Eq.~\eqn{eq:delta_P}
implies the strong coupling determination quoted in Eq.~\eqn{eq:alpha-result}.

\section{Discussion on previous $\cO(\alpha_s^4)$ determinations of $\alpha_s(m_\tau^2)$}

The integrals $A^{(n)}(\alpha_s)$ can be expanded in powers of $a_\tau\equiv\alpha_s(m_\tau^2)/\pi$,
$A^{(n)}(\alpha_s) = a_\tau^n + \cO(a_\tau^{n+1})$. One recovers in this way the naive perturbative expansion
$\delta_P\, =\, \sum_{n=1}\,  (K_n + g_n) \, a_\tau^n \,\equiv\,
\sum_{n=1}\,  r_n \, a_\tau^n $ \cite{LDP:92a}.
This approximation is known as {\it fixed-order perturbation theory} (FOPT), while
the improved expression \eqn{eq:r_k_exp}, keeping the non-expanded values for the integrals $A^{(n)}(\alpha_s)$,
is usually called {\it contour-improved perturbation theory} (CIPT) \cite{LDP:92a,PI:92}.
FOPT gives rise to a pathological non-convergent series (its radius of convergence is slightly smaller than the physical
value of $a_\tau$), because the long running of $\alpha_s(s)$ along the circle $|s|=m_\tau^2$ generates very large $g_n$ coefficients,
which depend on $K_{m<n}$ and $\beta_{m<n}$ \cite{LDP:92a}:
$g_1=0$, $g_2 =  3.56$, $g_3 = 19.99$, $g_4 = 78.00$, $g_5 = 307.78$. These corrections are much larger than the
original $K_n$ contributions and lead to values of $\alpha_s(m_\tau^2)$ smaller than \eqn{eq:alpha-result}.
FOPT suffers from a huge renormalization-scale dependence \cite{LDP:92a}. As shown in a recent detailed analysis \cite{ME:09},
the actual FOPT uncertainties are much larger than usually estimated. Once this is taken properly into account,
the FOPT results are consistent with CIPT, but their huge uncertainties make them irrelevant.

An ad-hoc model of higher-order coefficients for the Adler function has been recently advocated \cite{BJ:08}.
The model mixes three different types of renormalons ($n=-1$, 2 and 3) plus a linear polynomial, trying to enforce
a cancelation of the $K_n$ and $g_n$ coefficients in order to get a better behaviour of the FOPT series.
It contains 5 free parameters which are determined by the known coefficients of the Adler series and
the chosen value of $K_5$. Making a Borel summation
of the full renormalon series (BSR), one gets a sizeable positive contribution to $\delta_P$ from higher orders, implying
a smaller value for $\alpha_s(m_\tau^2)$ independently of the adopted FOPT \cite{BJ:08} or
CIPT \cite{CF:09} procedure. This model constitutes an interesting example of possible higher-order corrections,
making apparent that the associated uncertainties have to be carefully estimated. However, it cannot be used
to determine the physical value of $\alpha_s$, because the result is model dependent. Different assumptions about
the unknown $K_n$ coefficients would lead to different central values for $\alpha_s$.

The determination of $\alpha_s(m_\tau^2)$ in Eq.~\eqn{eq:alpha-result} agrees with Refs.~\cite{DHZ:05,ME:09}
and with the CIPT value in Ref.~\cite{BChK:08}. It is somewhat larger than the (CIPT) result extracted in Ref.~\cite{MY:08}
from pinched-weight moments (PWM) of the hadronic distribution, which seems to correspond to a very different value of $\delta_{NP}$.
After evolution up to the scale $M_Z$ \cite{Rodrigo:1998zd}, the strong coupling in Eq.~\eqn{eq:alpha-result} decreases to
\be \alpha_s(M_Z^2)\, =\, 0.1213\pm 0.0014\, ,\ee
in excellent agreement with the direct measurements at the Z peak.

\section*{Acknowledgements}
This work has been supported in part by
the EU Contract MRTN-CT-2006-035482 (FLAVIAnet),
by MICINN, Spain (grants FPA2007-60323 and Consolider-Ingenio 2010 CSD2007-00042 --CPAN--)
and by Generalitat Valenciana (Prometeo/2008/069, ACOMP/2009/312).


\begin{thebibliography}{10}

\bibitem{BNP:92}
 E. Braaten, S. Narison and A. Pich, Nucl. Phys. B 373 (1992) 581.

\bibitem{LDP:92a} F. Le Diberder and A. Pich, Phys. Lett. B 286 (1992) 147.

\bibitem{NP:88} S. Narison and A. Pich, Phys. Lett. B 211 (1988) 183.

\bibitem{BR:88} E. Braaten, Phys. Rev. Lett. 60 (1988) 1606;
           Phys. Rev. D 39 (1989) 1458.

\bibitem{QCD:94} A. Pich, Nucl. Phys. B (Proc. Suppl.) 39B,C (1995) 326;   
   arXiv:hep-ph/9701305.

\bibitem{DHZ:05} M. Davier, A. H\"ocker and Z. Zhang,
Rev. Mod. Phys. 78 (2006) 1043; 

\bibitem{QCD08} A. Pich, Nucl. Phys. B (Proc. Suppl.) 186 (2009) 187.

\bibitem{BChK:08} P.A. Baikov, K.G. Chetyrkin and J.H. K\"uhn, 
 Phys. Rev. Lett. 101 (2008) 012002.

\bibitem{DDMHZ:08} M. Davier et al.,
Eur. Phys. J. C 56 (2008) 305. 

\bibitem{BJ:08}
M. Beneke and M. Jamin, JHEP 0809 (2008) 044.

\bibitem{MY:08}
K.~Maltman and T. Yavin, Phys. Rev. D 78 (2008) 094020. 

\bibitem{ME:09} S. Menke, arXiv:0904.1796 [hep-ph].

\bibitem{CF:09} I. Caprini and J. Fischer, Eur. Phys. J. C 64 (2009) 35.  

\bibitem{CGP:08} O. Cata, M. Golterman and S. Peris, Phys. Rev. D 77 (2008) 093006. 

\bibitem{MS:88} W.J. Marciano and A. Sirlin, Phys. Rev. Lett. 61 (1988) 1815.

\bibitem{BL:90} E. Braaten and C.S. Li, Phys. Rev. D 42 (1990) 3888.

\bibitem{ER:02} J. Erler, Rev. Mex. Phys. 50 (2004) 200. 

\bibitem{PP:99} A. Pich and J. Prades,
 JHEP 9910 (1999) 004; 
    9806 (1998) 013. 

\bibitem{BChK:05}
P.A. Baikov, K.G. Chetyrkin and J.H.~K\"uhn, Phys. Rev. Lett. 95 (2005)
012003. 

\bibitem{LDP:92b} F. Le Diberder and A. Pich, Phys. Lett. B 289 (1992) 165.

\bibitem{ALEPH:05}
 ALEPH Collaboration, Phys. Rep. 421 (2005) 191; 
 Eur. Phys. J. C 4 (1998) 409; Phys. Lett. B 307 (1993) 209.

\bibitem{CLEO:95} CLEO Collaboration, Phys. Lett. B 356 (1995) 580.

\bibitem{OPAL:98} OPAL Collaboration, Eur. Phys. J. C 7 (1999) 571. 

\bibitem{PDG:08}
C. Amsler et al.,  {\em The Review of Particle Physics}, Phys. Lett. B 667 (2008) 1.

\bibitem{ChKT:79}
 K.G. Chetyrkin, A.L. Kataev and F.V. Tkachov, Phys. Lett.  85B (1979) 277.

\bibitem{DS:79}
 M. Dine and J. Sapirstein, Phys. Rev. Lett. 43 (1979) 668.

\bibitem{CG:80}
 W. Celmaster and R. Gonsalves, Phys. Rev. Lett. 44 (1980) 560.

\refjl{GKL:91}{S.G. Gorishny, A.L. Kataev and S.A. Larin}{Phys. Lett.}{B 259}{1991}{144}

\refjl{SS:91}{L.R. Surguladze and M.A. Samuel}{Phys. Rev. Lett.}{66}{1991}{560}

\bibitem{PI:92} A.A. Pivovarov, Z. Phys. C 53 (1992) 461.

\bibitem{Rodrigo:1998zd}
  G.~Rodrigo, A.~Pich and A.~Santamaria, Phys. Lett. B 424 (1998) 367.

\end{thebibliography}
\end{document}